\documentclass[12pt]{article}
\usepackage{epsfig}

\input paperdef

\newcommand\pubnumber{BNL--HET-01/7\\ CERN--TH/2001--056}
\newcommand\pubdate{\today}
\newcommand\hepnumber{hep-ph/0102317}

\def\csumb{$^*$HET, Brookhaven Natl.\ Lab., Upton NY 11973, USA\\
$^{\dagger}$CERN, TH Division, CH-1211 Geneva 23, Switzerland}

\def\Title#1{\begin{center} {\Large\bf #1 } \end{center}}
\def\Author#1{\begin{center}{ \sc #1} \end{center}}
\def\Address#1{\begin{center}{ \it #1} \end{center}}

\newcommand\pubblock{\rightline{\begin{tabular}{l} \pubnumber\\
         \pubdate\\ \hepnumber \end{tabular}}}
\newenvironment{Abstract}{\begin{quotation}  }{\end{quotation}}
\newenvironment{Presented}{\begin{quotation} \begin{center} 
             Presented by S.~Heinemeyer at the\end{center}
      \begin{center}\begin{large}}{\end{large}\end{center} \end{quotation}}
\def\Acknowledgments{\bigskip  \bigskip \begin{center}
          \large\bf Acknowledgments\end{center}}

\makeatletter
\def\section{\@startsection{section}{0}{\z@}{5.5ex plus .5ex minus
 1.5ex}{2.3ex plus .2ex}{\large\bf}}
\def\subsection{\@startsection{subsection}{1}{\z@}{3.5ex plus .5ex minus
 1.5ex}{1.3ex plus .2ex}{\normalsize\bf}}
\def\subsubsection{\@startsection{subsubsection}{2}{\z@}{-3.5ex plus
-1ex minus  -.2ex}{2.3ex plus .2ex}{\normalsize\sl}}

\renewcommand{\@makecaption}[2]{%
   \vskip 10pt
   \setbox\@tempboxa\hbox{\small #1: #2}
   \ifdim \wd\@tempboxa >\hsize     
       \small #1: #2\par          
     \else                        
       \hbox to\hsize{\hfil\box\@tempboxa\hfil}
   \fi}

 \def\citenum#1{{\def\@cite##1##2{##1}\cite{#1}}}
 
\newcount\@tempcntc
\def\@citex[#1]#2{\if@filesw\immediate\write\@auxout{\string\citation{#2}}\fi
  \@tempcnta\z@\@tempcntb\m@ne\def\@citea{}\@cite{\@for\@citeb:=#2\do
    {\@ifundefined
       {b@\@citeb}{\@citeo\@tempcntb\m@ne\@citea\def\@citea{,}{\bf ?}\@warning
       {Citation `\@citeb' on page \thepage \space undefined}}%
    {\setbox\z@\hbox{\global\@tempcntc0\csname b@\@citeb\endcsname\relax}%
     \ifnum\@tempcntc=\z@ \@citeo\@tempcntb\m@ne
       \@citea\def\@citea{,}\hbox{\csname b@\@citeb\endcsname}%
     \else
      \advance\@tempcntb\@ne
      \ifnum\@tempcntb=\@tempcntc
      \else\advance\@tempcntb\m@ne\@citeo
      \@tempcnta\@tempcntc\@tempcntb\@tempcntc\fi\fi}}\@citeo}{#1}}
\def\@citeo{\ifnum\@tempcnta>\@tempcntb\else\@citea\def\@citea{,}%
  \ifnum\@tempcnta=\@tempcntb\the\@tempcnta\else
  {\advance\@tempcnta\@ne\ifnum\@tempcnta=\@tempcntb \else\def\@citea{--}\fi
    \advance\@tempcnta\m@ne\the\@tempcnta\@citea\the\@tempcntb}\fi\fi}
\makeatother

\def\beq{\begin{equation}}
\def\eeq#1{\label{#1}\end{equation}}
\def\eeqn{\end{equation}}

\newenvironment{Eqnarray}%
   {\arraycolsep 0.14em\begin{eqnarray}}{\end{eqnarray}}
\def\beqa{\begin{Eqnarray}}
\def\eeqa#1{\label{#1}\end{Eqnarray}}
\def\eeqan{\end{Eqnarray}}

\let\bar=\overbar

\def\Dslash{\not{\hbox{\kern-4pt $D$}}}
\def\dslash{\not{\hbox{\kern-2pt $\del$}}}

\def\mt{m_t}

\def\mh{m_h}

\def\msb{{\bar{\ssstyle M \kern -1pt S}}}

\def\lsim{\mathrel{\raise.3ex\hbox{$<$\kern-.75em\lower1ex\hbox{$\sim$}}}}
\def\gsim{\mathrel{\raise.3ex\hbox{$>$\kern-.75em\lower1ex\hbox{$\sim$}}}}

\begin{document}
\begin{titlepage}
\pubblock

\vfill
\def\thefootnote{\fnsymbol{footnote}}
\Title{Precision Observables in the MSSM:\\[0.5em]
Leading Electroweak Two-loop Corrections}
\vfill
\Author{S.~Heinemeyer$^*$ and G.~Weiglein$^\dagger$} 
\Address{\csumb}
\vfill
\begin{Abstract}
The leading electroweak MSSM \twol\ corrections to 
the $\rho$-parameter are calculated. They are
obtained by evaluating the \twol\ self-energies of the $Z$~and the
$W$~boson at \order{\gf^2\mt^4} 
in the limit of heavy scalar \mbox{quarks}.
A very compact expression is derived, depending on the ratio of the
$\cp$-odd Higgs boson mass, $\MA$, and the top quark mass,
$\mt$. Expressions for the limiting cases $\MA \gg \mt$ and $\MA \ll \mt$
are also given. The decoupling of the non-SM contribution in the limit
$\MA \to \infty$ is verified at the \twol\ level.
The numerical effect of the leading electroweak MSSM \twol\ corrections
is analyzed in comparison with the leading corrections of
\order{\gf^2\mt^4} in the SM and with the $\oaas$ corrections
in the MSSM.
\end{Abstract}
\vfill
\begin{Presented}
5th International Symposium on Radiative Corrections \\ 
(RADCOR--2000) \\[4pt]
Carmel CA, USA, 11--15 September, 2000
\end{Presented}
\vfill
\end{titlepage}
\def\thefootnote{\arabic{footnote}}
\setcounter{footnote}{0}
%


\section{Introduction}
\label{sec:intro}

Theories based on Supersymmetry (SUSY) \cite{susy} 
are widely considered as the
theoretically most appealing extension of the Standard Model (SM). 
They predict the existence of scalar partners $\tilde{f}_L, 
\tilde{f}_R$ to each SM chiral fermion, and spin--1/2 partners to the 
gauge bosons and to the scalar Higgs bosons. So far, the direct search for 
SUSY particles has not been successful. 
One can only set lower bounds of ${\cal O}(100)$ GeV on 
their masses~\cite{pdg}. 
Contrary to the SM, two Higgs doublets are
required in the Minimal Supersymmetric Standard Model (MSSM) resulting
in five physical Higgs bosons~\cite{hhg}. The 
direct search resulted in lower limits of about $90 \gev$ for the
neutral Higgs bosons~\cite{lephiggs}. 

An alternative way to probe SUSY is to search for the virtual effects of the 
additional particles via precision observables. 
The most prominent role in this respect plays the
$\rho$-parameter~\cite{rho}. The leading radiative corrections to the
$\rho$-parameter, $\De\rho$, 
constitute the leading process-independent corrections
to many electroweak precision observables, such as the
$W$~boson mass, $\MW$, and the effective leptonic weak mixing angle,
$\sweff$. Within the MSSM
the full \onel\ corrections to $\MW$ and $\sweff$ have been calculated
already several years ago~\cite{dr1lA,dr1lB}.
More recently also the leading \twol\ corrections of $\oaas$ to the
quark and scalar quark loops for $\De\rho$ 
and $\MW$ have been obtained~\cite{dr2lA,dr2lB}. Contrary to the SM
case, these \twol\ corrections turned out to increase the \onel\
contributions, leading to an enhancement of the latter 
of up to 35\%~\cite{dr2lA}.

We summarize here the result for the leading \twol\ corrections to 
$\De\rho$ at \order{\gf^2\mt^4}~\cite{dr2lal2}. 
For a large SUSY scale, $\msusy \gg \MZ$, the SUSY
contributions decouple from physical observables. This has been verified
with existing results at the
\onel~\cite{decoupling1l} and at the \twol\ level~\cite{dr2lA,dr2lal2}.
Therefore, in the case of large $\msusy$ the leading electroweak \twol\ 
corrections in the MSSM are obtained in the limit where besides the SM 
particles only the two Higgs doublets needed in the MSSM are active.
We derive the result for the \order{\gf^2\mt^4}~\cite{dr2lal2}
corrections in this case and provide a compact analytical formula for
it, depending on the $\cp$-odd Higgs boson mass, $\MA$, and the top quark 
mass, $\mt$. Furthermore, we present formulas for the limiting cases 
$\MA \gg \mt$ (i.e.\ the SM limit) and $\MA \ll \mt$.
The numerical effect of the \order{\gf^2\mt^4} corrections is
compared with the corresponding SM result~\cite{drSMgf2mt4}
and the gluon-exchange correction of $\oaas$ in the MSSM.


\newpage
\section{Calculation of the \order{\gf^2\mt^4} corrections}
\label{sec:calc}

\subsection{$\De\rho$ and the Higgs sector}
\label{subsec:mssmhiggs}

The quantity $\De\rho$, 
\BE
\De\rho = \frac{\Si_Z(0)}{\MZ^2} - \frac{\Si_W(0)}{\MW^2} ,
\label{delrho}
\EE
 parameterizes the leading universal corrections to the electroweak
precision observables induced by the
mass splitting between fields in an isospin doublet~\cite{rho}.
$\Si_{Z,W}(0)$ denote the transverse parts of the unrenormalized $Z$~and 
$W$~boson self-energies at zero momentum transfer, respectively.
The shifts induced by $\De\rho$ in the prediction for the $W$ boson
mass, $\MW$, and the effective leptonic weak mixing angle, $\sweff$, are
approximately given by
\BE
\de\MW \approx \frac{\MW}{2}\frac{\cw^2}{\cw^2 - \sw^2} \De\rho, \quad
\de\sweff \approx - \frac{\cw^2 \sw^2}{\cw^2 - \sw^2} \De\rho .
\label{precobs}
\EE


\bigskip
Contrary to the SM, in the MSSM two Higgs doublets
are required~\cite{hhg}.
At the tree-level, the Higgs sector can be described in terms of two  
independent parameters (besides $g$ and $g'$): the ratio of the two
vacuum expectation values,  
$\tb = v_2/v_1$, and $M_A$, the mass of the $\cp$-odd $A$ boson.
The diagonalization of the bilinear part of the Higgs potential,
i.e.\ the Higgs mass matrices, is performed via orthogonal
transformations with the angle $\al$ for the $\cp$-even part and with
the angle $\be$ for the $\cp$-odd and the charged part.
The mixing angle $\al$ is determined at lowest order through
\BE
\tan 2\al = \tan 2\be \; \frac{\MA^2 + M_Z^2}{\MA^2 - M_Z^2} ;
\qquad  -\frac{\pi}{2} < \al < 0~.
\label{alphaborn}
\EE
One gets the following Higgs spectrum:
\BEA
\mbox{2 neutral bosons},\, {\cal CP} = +1 &:& h^0, H^0 \non \\
\mbox{1 neutral boson},\, {\cal CP} = -1  &:& A^0 \non \\
\mbox{2 charged bosons}                   &:& H^+, H^- \non \\
\mbox{3 unphysical scalars}      &:& G^0, G^+, G^- .
\EEA
The tree-level masses, expressed through $\MZ, \MW$ and $\MA$, 
are given by
\BEA
\mh^2 &=& \edz \KKL \MA^2 + \MZ^2 - 
          \sqrt{(\MA^2 + \MZ^2)^2 - 4 \MA^2\MZ^2 \CQZb} \KKR \non \\
\mH^2 &=& \edz \KKL \MA^2 + \MZ^2 + 
          \sqrt{(\MA^2 + \MZ^2)^2 - 4 \MA^2\MZ^2 \CQZb} \KKR \non \\
\mHp^2 &=& \MA^2 + \MW^2 \non \\
\mG^2 &=& \MZ^2 \non \\
\mGp^2 &=& \MW^2 ,
\label{allhiggsmasses}
\EEA
where the last two relations, which assign mass parameters to the
unphysical scalars $G^0$ and $G^{\pm}$, are to be understood in the
Feynman gauge.


\subsection{Evaluation of the \order{\gf^2\mt^4} contributions}
\label{subsec:gf2mt4eval}

\begin{figure}[htb!]
\begin{center}
\mbox{
\psfig{figure=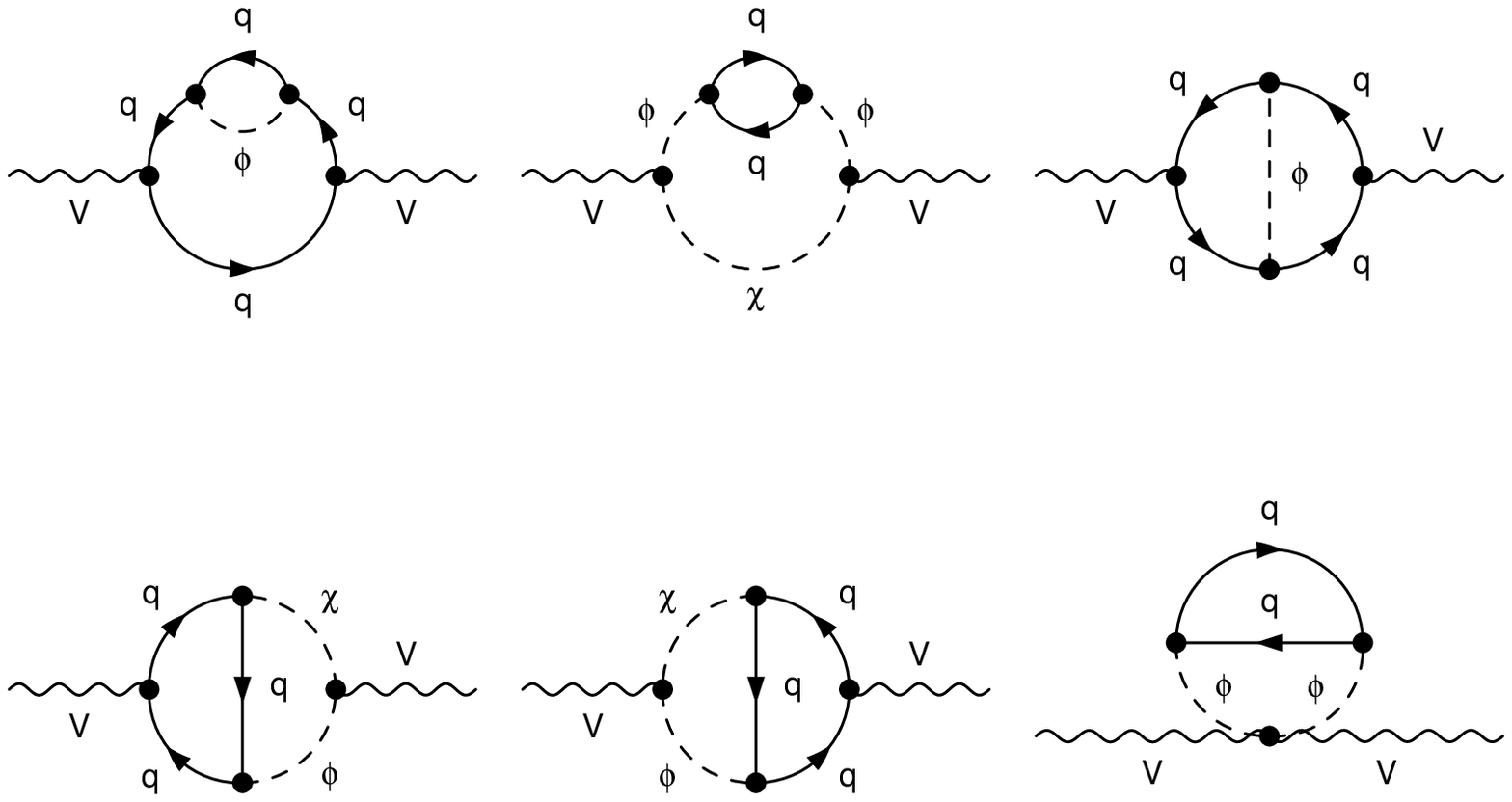,width=8cm,bbllx=140pt,bblly=360pt,
                                          bburx=460pt,bbury=580pt}}
\end{center}
\caption[]{
Generic Feynman diagrams for the vector boson self-energies\\ 
$(V = \{Z,W\}, q = \{t, b\}, \phi,\chi = \{h, H, A, H^\pm, G, G^\pm\})$. 
}
\label{fig:fdvb2l}
\end{figure}

In order to calculate the \order{\gf^2\mt^4} corrections to $\De\rho$
in the approximation that all superpartners are heavy so that their
contribution decouples,
the Feynman diagrams generically depicted in
\reffi{fig:fdvb2l} have to be evaluated for the $Z$ boson ($V = Z$)
and the $W$ boson ($V = W$) self-energy. We have taken into account
all possible combinations of the $t/b$ doublet and the full Higgs
sector of the MSSM, see \refse{subsec:mssmhiggs}. 

\begin{figure}[htb!]
\vspace{1em}
\begin{center}
\mbox{
\psfig{figure=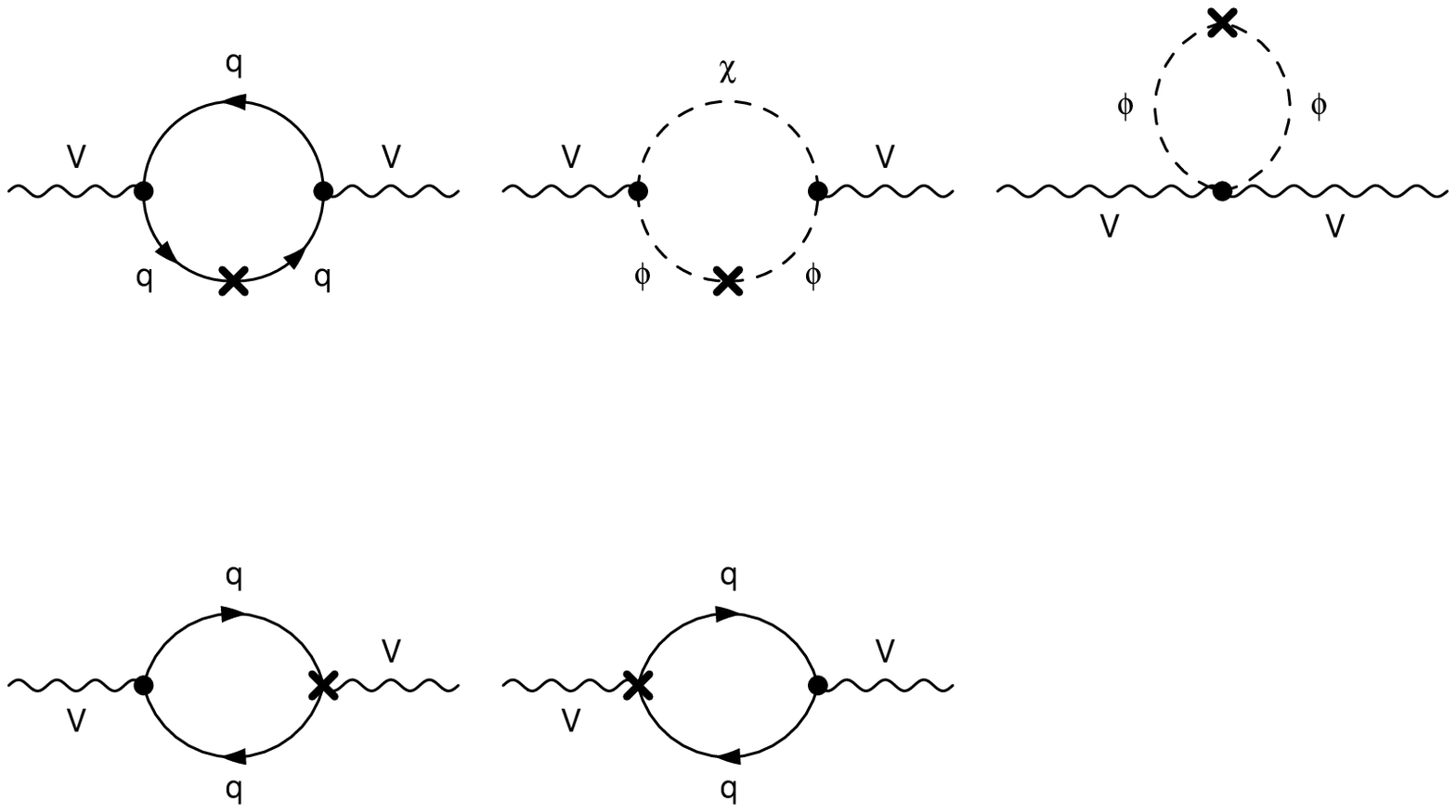,width=8cm,bbllx=130pt,bblly=360pt,
                                          bburx=450pt,bbury=580pt}}
\end{center}
\caption[]{
Generic Feynman diagrams for the vector boson self-energies with
counter term insertion 
$(V = \{Z,W\}, q = \{t, b\}, \phi,\chi = \{h, H, A, H^\pm, G, G^\pm\})$. 
}
\label{fig:fdvb1lct}
\end{figure}

\begin{figure}[htb!]
\vspace{1em}
\vspace{2em}
\begin{center}
\mbox{
\psfig{figure=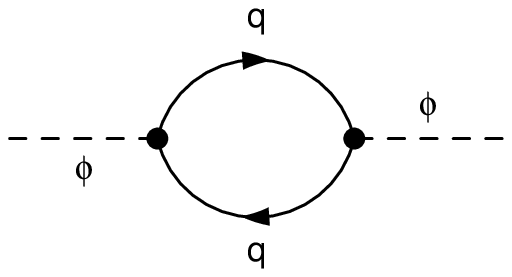,width=3cm,bbllx=100pt,bblly=460pt,
                                         bburx=220pt,bbury=550pt}}
\hspace{3em}
\mbox{
\psfig{figure=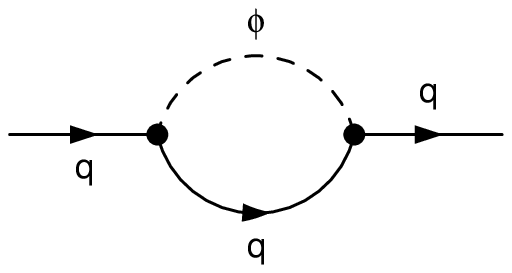,width=3cm,bbllx=100pt,bblly=460pt,
                                         bburx=220pt,bbury=550pt}}
\end{center}
\vspace{-3em}
\caption[]{
Generic Feynman diagrams for the counter term insertions\\
$(q = \{t, b\}, \phi = \{h, H, A, H^\pm, G, G^\pm\})$. 
}
\label{fig:fdcti}
\end{figure}

The \twol\ diagrams shown in \reffi{fig:fdvb2l} have to be
supplemented with the corresponding \onel\ diagrams with subloop
renormalization, depicted generically in \reffi{fig:fdvb1lct}. The
corresponding insertions for the fermion and Higgs mass
counter terms are shown in \reffi{fig:fdcti}. 

The amplitudes of all Feynman diagrams, shown in 
\reffis{fig:fdvb2l}--\ref{fig:fdcti}, have been created with the
program {\em FeynArts2.2}~\cite{feynarts}, making use of a recently
completed model file for the MSSM%
\footnote{
Only the non-SM like counter terms had to be added. 
}%
.
The algebraic evaluation and reduction to scalar integrals has been 
performed with the program \tc~\cite{2lred}. 
(Further details about the evaluations with {\em FeynArts2.2} and
\tc\ can be found in \citere{SHacat2000}.) 
As a result we
obtained the analytical expression for $\De\rho$ depending on the
\onel\ functions $A_0$ and $B_0$~\cite{a0b0c0d0} and on the \twol\
function $T_{134}$~\cite{2lred,t134}. For the further evaluation the
analytical expressions for $A_0$, $B_0$ and $T_{134}$ have been
inserted. In order to derive the leading contributions of
\order{\gf^2\mt^4} we extracted a prefactor $h_t^4 \sim \gf^2\mt^4$.
Its coefficient can be evaluated in the limit where $\MW$ and $\MZ$ 
(and also $\mb$) are
set to zero. Furthermore we made use of the mass relations in the
MSSM Higgs sector, see \refeq{allhiggsmasses}. In the limit 
$\MW, \MZ \to 0$ they reduce to
\BEA
\mh^2 &=& 0 \non \\
\mH^2 &=& \MA^2 \non \\
\mHp^2 &=& \MA^2 \non \\
\mG^2 &=& 0 \non \\
\mGp^2 &=& 0 .
\label{allhiggsmassesmw0}
\EEA
In the limit $\MZ \to 0$ the relation between the angles $\al$ and
$\be$, see \refeq{alphaborn}, becomes very simple, 
$\al = \be - \pi/2$, i.e. $\Sa = -\Cb$, $\Ca = \Sb$.
The coefficient of the leading \order{\gf^2\mt^4} term thus depends only
on the top quark mass, $\mt$, the $\cp$-odd
Higgs boson mass, $\MA$, and $\tb$ (or $\sbe = \tb/\sqrt{1 + \TQb}$). 

%
%

We explicitly verified the UV-finiteness of our result. As a further 
consistency check of our method we also recalculated the SM result 
for the \order{\gf^2\mt^4} corrections with arbitrary values of the
Higgs boson mass, as given in \citere{drSMgf2mt4MH}, and found perfect 
agreement.


\section{Analytical results}

\subsection{The full result}

The analytical result obtained as described in
\refse{subsec:gf2mt4eval} can conveniently be expressed in terms of 
\BE
\label{defa}
a \equiv \frac{\mt^2}{\MA^2} .
\EE
\newcommand{\sqa}{\sqrt{1 - 4\,a}}

\noindent
The corresponding \twol\ contribution to $\De\rho$ then reads:
\BEA
\De\rho_{1,\hi}^{\SU} &=& 3 \, \frac{\gf^2}{128 \,\pi^4} \, \mt^4 \, 
                      \frac{1 - \sbe^2}{\sbe^2 \, a^2} \times \non \\
&& \Bigg\{ 
   \Liz{ \KL 1 - \sqa \KR/2} \frac{8}{\sqa} \La \non \\
&& - 2\, \Liz{1 - \ed{a}} \KKL 5 - 14 a + 6 a^2 \KKR \non \\
&& + \log^2(a) \KKL 1 + \frac{2}{\sqa} \La \KKR \;
   - \log(a) \Big[ 2 - 20 \, a \Big] \non \\
&& - \log^2 \KL \frac{1 - \sqa}{2} \KR \frac{4}{\sqa} \La \non \\ 
&& + \log \KL \frac{1 - \sqa}{1 + \sqa} \KR 
     \sqa (1 - 2 \, a) \non \\
&& - \log\KL |1/a - 1| \KR \, (a - 1)^2 \non \\
&& + \pi^2 \KKL \frac{2\sqa}{-3 + 12 \, a} \La
                + \ed{3} - 2\, a^2 \frac{\sbe^2}{1 - \sbe^2} \KKR 
   - 17 a + 19 \frac{a^2}{1 - \sbe^2}
   \Bigg\} ,
\label{drallma}
\EEA
with 
\BE
\La = 3 - 13\,a + 11\,a^2 .
\EE
In the limit of large $\tb$ (i.e.\ $(1 - \sbe^2) \ll 1$) one obtains
\BE
\De\rho_{1,\hi}^{\SU} = 3 \, \frac{\gf^2}{128 \,\pi^4} \, \mt^4 \, 
\KKL \frac{19}{\sbe^2} - 2\,\pi^2 + \orderm{1 - \sbe^2} \KKR .
\label{drallmalargetb}
\EE
Thus for large $\tb$ the SM limit with $\MH^{\SM} \to 0$~\cite{drSMgf2mt4} 
is reached.


\subsection{The expansion for large $\MA$}

The result for $\De\rho_{1,\hi}^{\SU}$ in \refeq{drallma}
can be expanded for small values of
$a$, i.e.\ for large values of $\MA$:
\BEA
\De\rho_{1,\hi}^{\SU} &=& 3 \, \frac{\gf^2}{128 \,\pi^4} \, \mt^4 \, 
                      \times \non \\
&& \Bigg\{
   19 - 2 \pi^2 \non \\
&& -\frac{1 - \sbe^2}{\sbe^2} \Bigg[ \KL \log^2 a + \frac{\pi^2}{3} \KR
           \KL 8 a + 32 a^2 + 132 a^3 + 532 a^4 \KR \non \\
\label{drlargema}
&& + \log(a) \ed{30} \KL 560 a + 2825 a^2 + 11394 a^3 + 45072 a^4 \KR 
     \non \\
&& - \ed{1800} \KL 2800 a + 66025 a^2 + 300438 a^3 + 1265984 a^4 \KR
     \non \\
&& + {\cal O}\KL a^5 \KR \Bigg] 
   \Bigg\} .
\EEA
In the limit $a \to 0$ one obtains
\BE
\De\rho_{1,\hi}^{\SU} = 3 \, \frac{\gf^2}{128 \,\pi^4} \, \mt^4 \, 
 \KKL 19 - 2\,\pi^2 \KKR + \cO(a) ,
\label{drSMlimit}
\EE
i.e.\ exactly the SM limit for $\MH^{\SM} \to 0$ is
reached. 
This constitutes an important consistency check: in the limit $a \to 0$ 
the heavy Higgs bosons decouple from the theory. Thus only the
lightest $\cp$-even Higgs boson remains, which has in the
\order{\gf^2\mt^4} approximation the mass $\mh = 0$, see
\refeq{allhiggsmassesmw0}. This decoupling of the non-SM contributions
in the limit where the new scale (i.e.\ in the present case $\MA$) is
made large is explicitly seen here at the two-loop level.



\subsection{The expansion for small $\MA$}

\newcommand{\ha}{\hat a}
The result for $\De\rho_{1,\hi}^{\SU}$ in \refeq{drallma}
can also be expanded for large values of
$a$, i.e.\ for small values of $\MA$ (with $\ha = 1/a$):
\BEA
\De\rho_{1,\hi}^{\SU} &=& 3 \, \frac{\gf^2}{128 \,\pi^4} \, \mt^4 \, 
                      \times \non \\
&& \Bigg\{
   \log^2(\ha) \, \ha^2 \KKL -1 + \ed{\sbe^2} \KKR \non \\
&& + \log(\ha) \frac{1 - \sbe^2}{210 \sbe^2} 
     \KKL -2100 \ha + 350 \ha^2 + 504 \ha^3 + 341 \ha^4 \KKR \non \\
&& + \pi^2 \frac{2}{3 \sbe^2}
     \KKL -3 + 7 \ha (1 - \sbe^2) - 2 \ha^2 (1 - \sbe^2) \KKR \non \\
&& - \pi \sqrt{\ha} \frac{1 - \sbe^2}{256 \sbe^2}
     \KKL 1024 - 640 \ha + 56 \ha^2 + 3 \ha^3 \KKR \\
&& + \frac{19}{\sbe^2} 
   - \frac{1 - \sbe^2}{22050 \sbe^2} 
     \KKL 970200 \ha - 376075 \ha^2 + 24843 \ha^3 + 6912 \ha^4 \KKR 
   + {\cal O} \KL \ha^5 \KR \Bigg\} . \non 
\label{drsmallma}
\EEA
In the limit $\ha \to 0$ or $a \to \infty$ one obtains
\BE
\De\rho_{1,\hi}^{\SU} = 3 \, \frac{\gf^2}{128 \,\pi^4} \, \mt^4 \, 
 \ed{\sbe^2} \KKL 19 - 2\,\pi^2 \KKR + \cO(\ha) .
\label{drlargetb}
\EE


\section{Numerical analysis}
\label{sec:numanal}

\subsection{The expansion formula}
\label{subsec:expansion}

\begin{figure}[htb!]
\begin{center}
\mbox{
\psfig{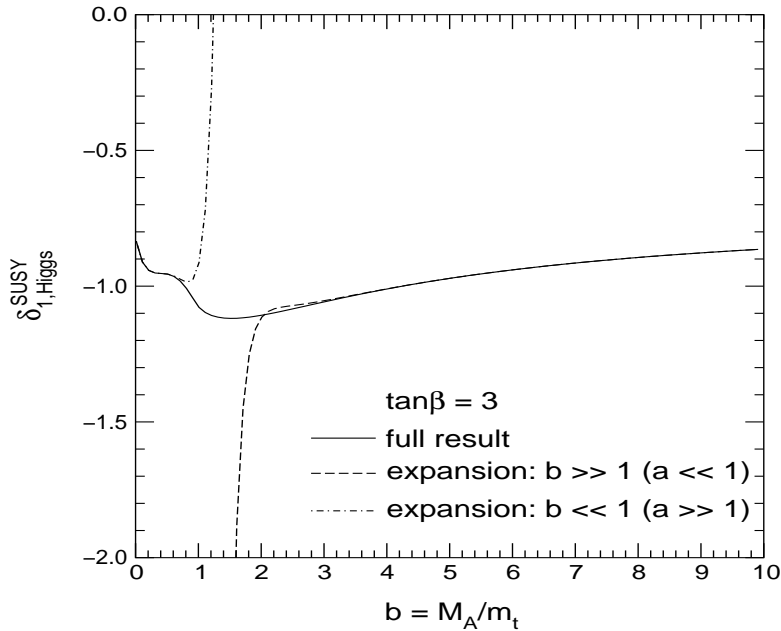}}
\end{center}
\caption[]{
The quality of the expansion formulas, \refeqs{drlargema} and
(\ref{drsmallma}), is shown as a function of 
$b = \MA/\mt \, (\equiv 1/\sqrt{a})$.
}
\label{fig:expansion}
\end{figure}

We first analyze the validity of the two expansion formulas,
\refeqs{drlargema} and (\ref{drsmallma}).
In \reffi{fig:expansion} we show the result for $\de_{1,\hi}^{\SU}$,
defined by
\BE
\De\rho_{1,\hi}^{\SU} = 3 \, \frac{\gf^2}{128\,\pi^4} \mt^4
                          \times \de_{1,\hi}^{\SU} ~,
\EE
as a function of $b = \MA/\mt \, (\equiv 1/\sqrt{a})$ for $\tb = 3$.
The expansion for $b \ll 1$ is sufficiently accurate nearly up to $b = 1$. 
The other expansion gives accurate results for $b \gsim 2$.
For larger $\tb$ the expansion becomes better, enlarging the validity
region for the large $\MA$ expansion up to $b \gsim 1$. 

\subsection{Effects on precision observables}
\label{subsec:precobs}

\begin{figure}[htb!]
\begin{center}
\mbox{
\epsfig{figure=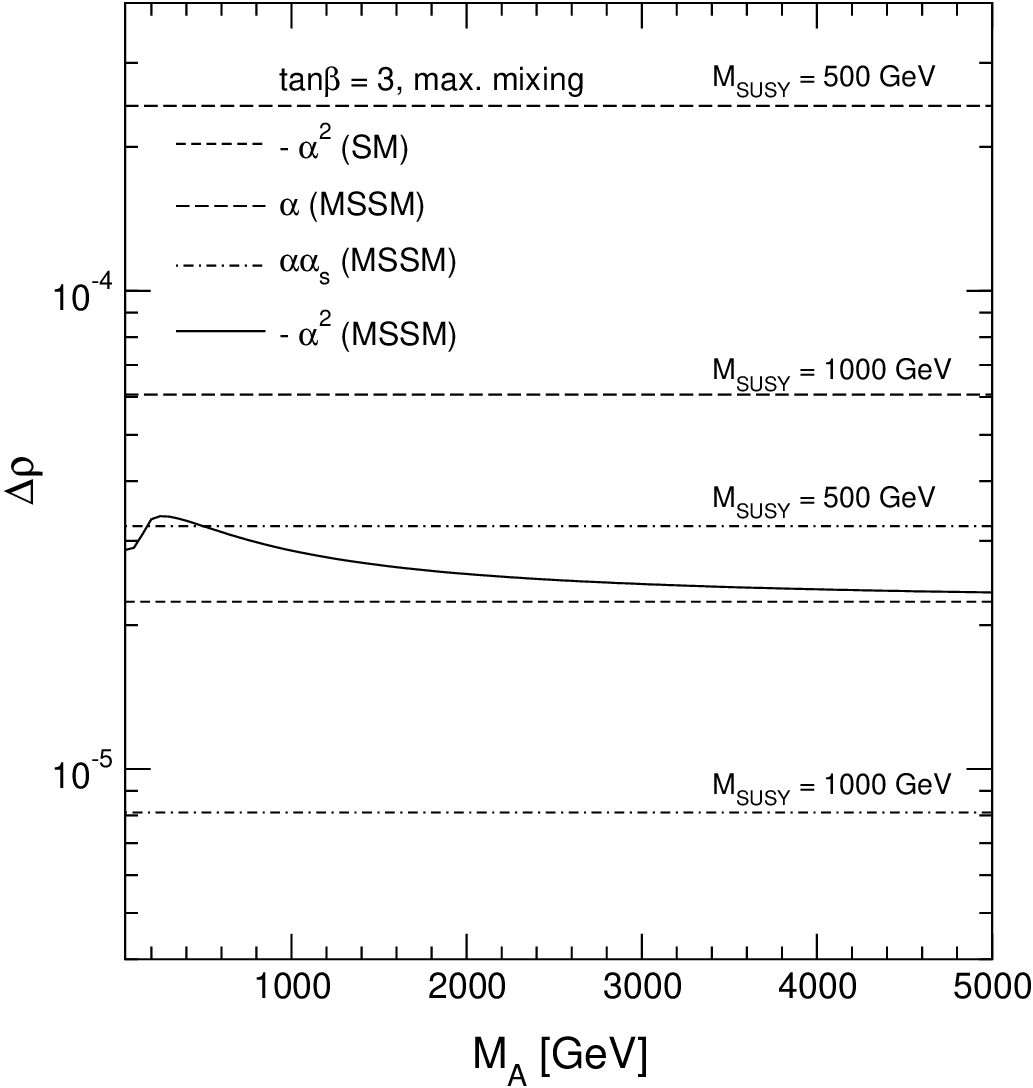,width=7cm,height=6.5cm}
\hspace{1em}
\epsfig{figure=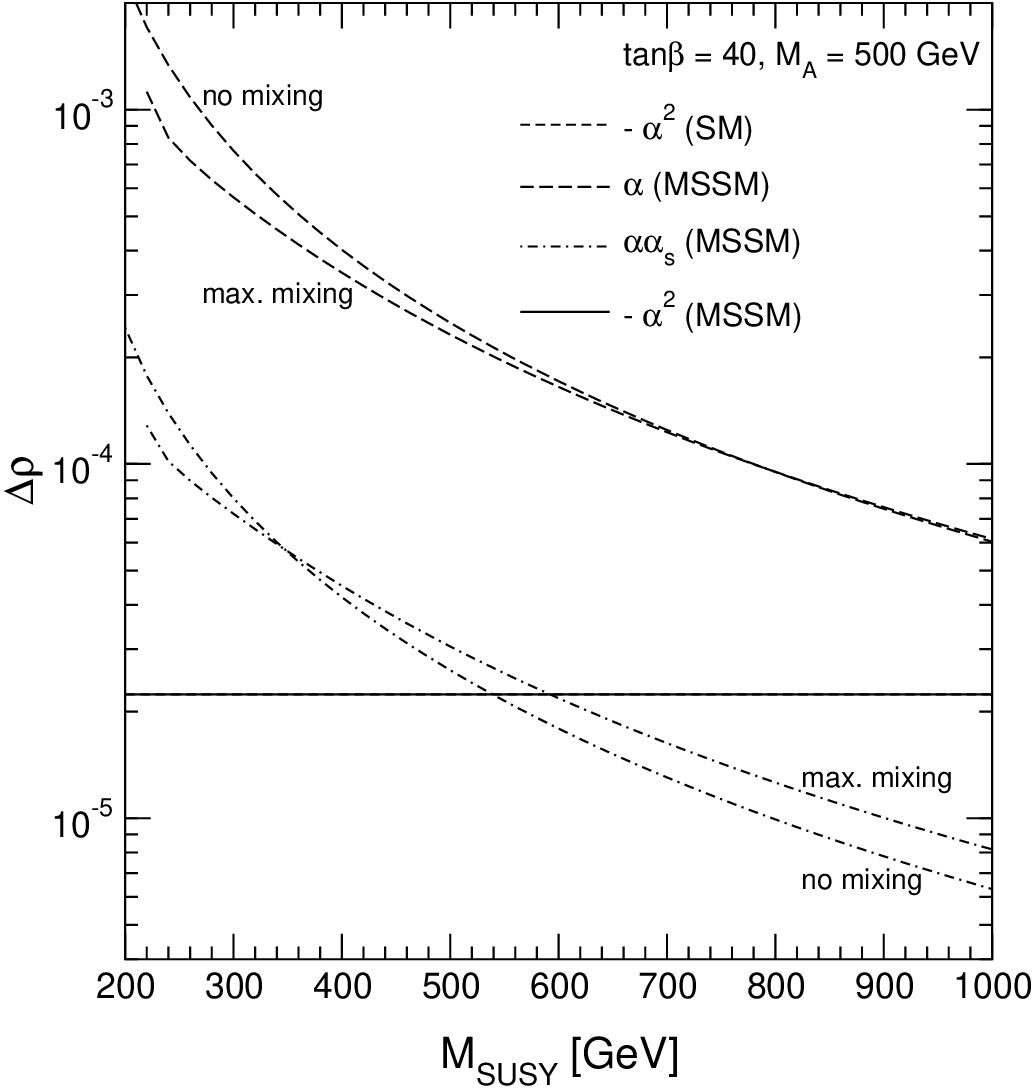,width=7cm,height=6.5cm}
}
\end{center}
\caption[]{
The contribution of the leading \order{\al^2} MSSM corrections,
$\De\rho_{1,\hi}^{\SU,\,\al^2}$, is shown as a function of $\MA$ for
$\tb = 3$ (left plot) and as a function of $\msusy$ for $\tb = 40$
(right plot).
In the left plot the case of maximal $\Stop$ mixing is shown, while the
right plot displays both the no-mixing and the maximal-mixing case.
$\De\rho_{1,\hi}^{\SU,\,\al^2}$ is compared with the leading
\order{\al^2} SM contribution with $\MH^{\SM} = 0$ and with the
leading MSSM corrections originating from the $\Stop/\Sbot$ sector of
\order{\al} and \order{\al\als}.
Both \order{\al^2} contributions are
negative and are for comparison shown with reversed sign.
}
\label{fig:delrho}
\end{figure}

In this section we analyze the
numerical effect on the precision observables $\MW$ and $\sweff$,
see \refeq{precobs}, induced by the additional contribution to $\De\rho$.
In \reffi{fig:delrho} the size of the leading \order{\al^2} MSSM corrections,
\refeq{drallma}, is compared for $\tb = 3, 40$ with the leading 
\order{\al^2} contribution
in the SM for $\MH^{\SM} = 0$~\cite{drSMgf2mt4}, with the leading MSSM
corrections arising from the $\Stop/\Sbot$ sector at
\order{\al}~\cite{dr1lB}, and with the corresponding gluon-exchange
contributions of \order{\al\als}~\cite{dr2lA} (the \order{\al\als}
gluino-exchange contributions~\cite{dr2lA}, which go to zero for large
$\mgl$, have been omitted here). For illustration, the left plot ($\tb =
3$) is shown as a function of $\MA$, which affects only the
\order{\al^2} MSSM contributions, while the right plot ($\tb = 40$) 
is given as a function of the common SUSY mass scale in the scalar quark
sector, $\msusy$, which affects only the \order{\al} and \order{\al\als}
MSSM contributions. We have furthermore chosen the case of ``maximal
mixing'' in the scalar top sector, which is realized by setting the 
off-diagonal term in the $\Stop$~mass matrix, $\Xt$, to $\Xt = 2
\msusy$ and yields the maximal value for $\mh$ for a given $\tb$
(see \citere{mhiggs135} for details). In the right plot the case of no
mixing, $\Xt = 0$, is also shown. The mixing in the scalar bottom
sector has been determined by using a bottom quark mass of 
$\mb = 4.5 \gev$, and by setting the trilinear couplings to $\Ab =
\At$ and the Higgs mixing parameter to $\mu = 200 \gev$. The
\order{\al^2} contributions in the SM and the MSSM are negative and are
for comparison shown with reversed sign.

While for small values of $\msusy$ the \order{\al\als} gluon-exchange
contribution in the MSSM is much larger than the \order{\al^2}
contribution from \refeq{drallma} (note that in this region of parameter
space the approximation of neglecting the scalar-quark contributions in
the \order{\al^2} result is no longer valid), they are of approximately
equal magnitude for $\msusy \approx 500 \gev$ (this refers to both the
no-mixing and the maximal-mixing case) and compensate each other
as they enter with different sign. In this region the two-loop
contributions are about one order of magnitude smaller than the 
\order{\al} MSSM contribution. For $\msusy = 1000 \gev$ the leading MSSM
\order{\al^2} contribution is about three times bigger than the
\order{\al\als} gluon-exchange contribution in the MSSM.

%
\begin{figure}[hb!]
\vspace{1em}
\begin{center}
\mbox{
\epsfig{figure=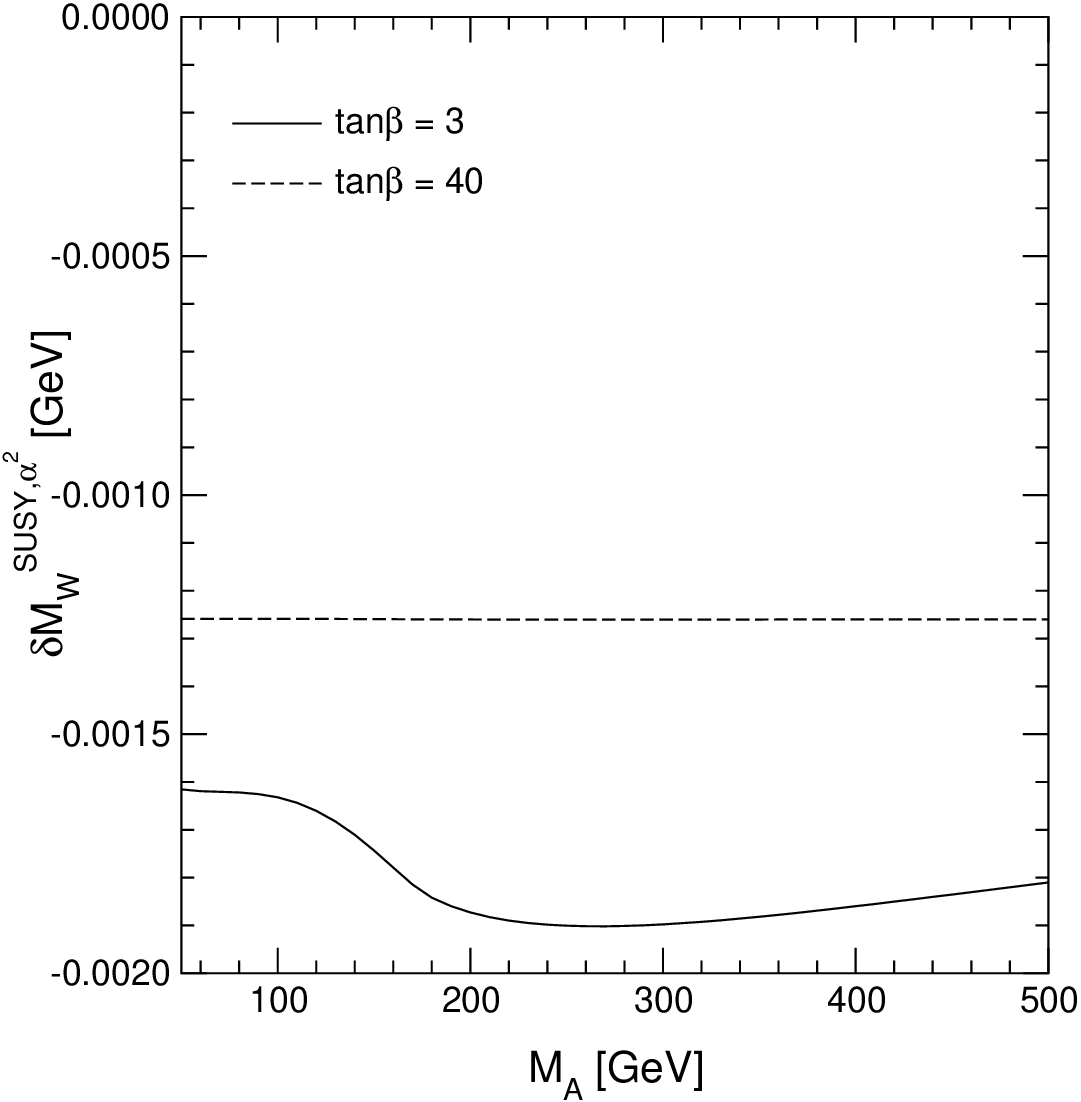,width=7cm,height=6.0cm}
\hspace{1em}
\epsfig{figure=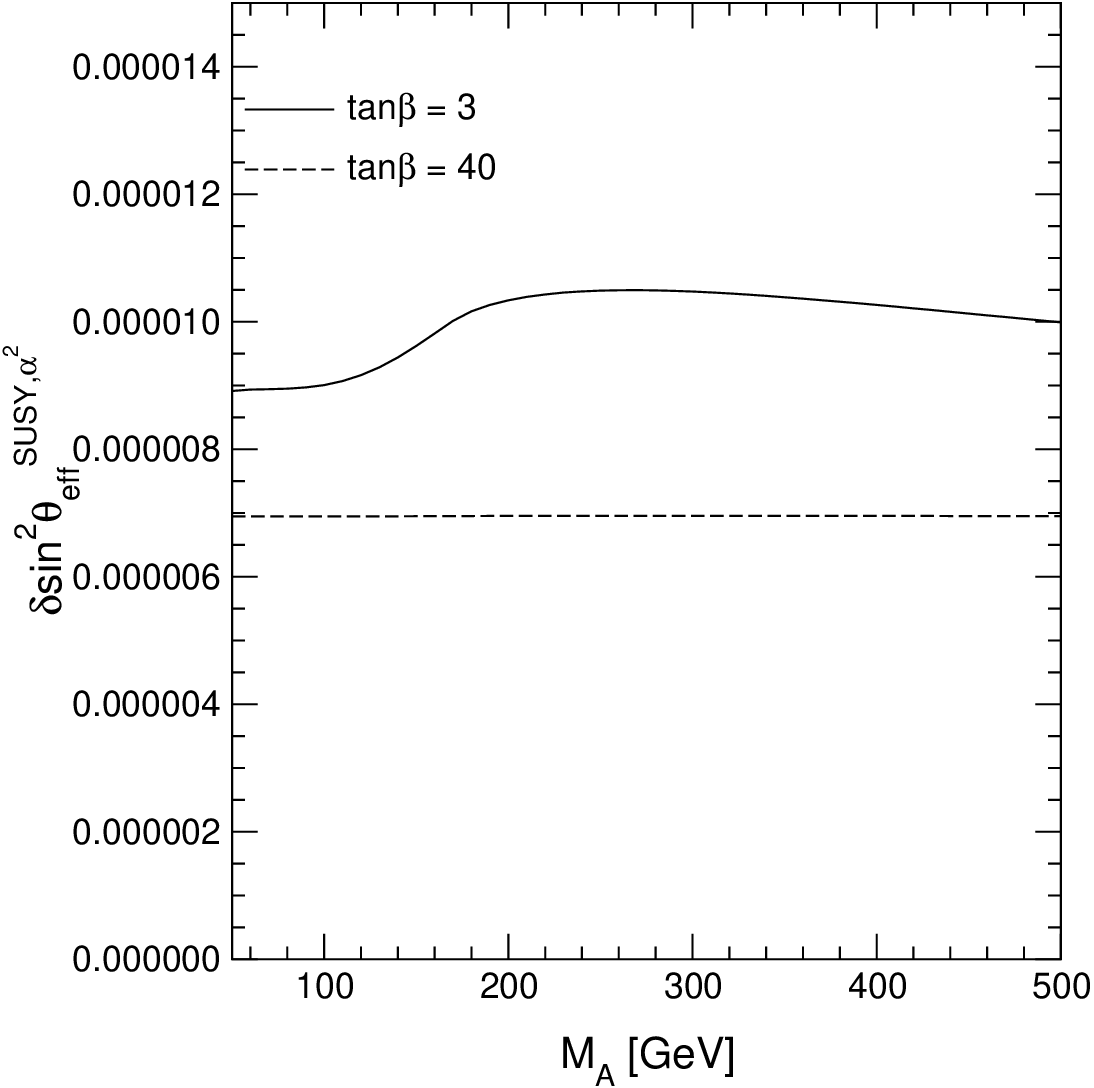,width=7cm,height=6.0cm}
}
\end{center}
\caption[]{
The leading \order{\al^2} MSSM contribution to $\de\MW$ (left plot) and
$\de\sweff$ (right plot) is shown as a function of $\MA$ for 
$\tb = 3, 40$.  
}
\label{fig:delPO}
\end{figure}

For small $\tb$ (left plot of \reffi{fig:delrho}) and moderate $\MA$
($\MA \approx 300 \gev$) the  
new \order{\al^2} MSSM corrections are about two times larger than the 
leading \order{\al^2} contributions in the SM for $\MH^{\SM} = 0$. For
large $\MA$ the decoupling of the extra contributions in the MSSM takes
place and the \order{\al^2} MSSM correction approaches the value of the 
leading \order{\al^2} contributions in the SM for $\MH^{\SM} = 0$, as
indicated in \refeqs{drlargema}, (\ref{drSMlimit}). For large $\tb$
(right plot of \reffi{fig:delrho}) the \order{\al^2} MSSM correction
and the \order{\al^2} 
contribution in the SM for $\MH^{\SM} = 0$ are indistinguishable in the
plot, in accordance with \refeq{drallmalargetb}.

It is well known that the \order{\al^2} SM result with $\MH^{\SM} = 0$
underestimates the result with realistic values of $\MH^{\SM}$ by about
one order of magnitude~\cite{drSMgf2mt4MH}. One can expect a similar effect 
in the MSSM once higher order corrections to the Higgs boson sector are 
properly taken into account, which can enhance $\mh$ up to 
$\mh \lsim 130 \gev$~\cite{mhiggs135}, see \citere{dr2lal2}. 

In \reffi{fig:delPO} the approximation formulas given in \refeq{precobs}
have been employed for determining the shift induced in $\MW$ and
$\sweff$ by the new \order{\al^2} correction to $\De\rho$.
In \reffi{fig:delPO} the effect for both precision observables is
shown as a fuction of $\MA$ for $\tb = 3, 40$. 
The effect on $\de\MW$ varies between $-1.5 \mev$ and $-2 \mev$ 
for small $\tb$ and
is almost constant, $\de\MW \approx -1.25 \mev$, for $\tb = 40$.
As above, the constant behavior can be
explained by the analytical decoupling of $\tb$ when $\tb \gg 1$, see
\refeq{drallmalargetb}. The induced shift in $\sweff$ lies at or below 
$1 \times 10^{-5}$ and shows the same qualitative $\tb$ dependence as 
$\de\MW$.


\section{Conclusions}

We have calculated the leading \order{\gf^2\mt^4} corrections to 
$\De\rho$ in the MSSM in the limit of heavy squarks. Short
analytical formulas have been obtained for the full result as well as
for the cases $\MA \gg \mt$ and $\MA \ll \mt$. 
As a consistency check we verified that from the MSSM result the
corresponding SM result can be obtained in the decoupling limit 
(i.e.\ $\MA \to \infty$). 

Numerically we compared the effect of the new contribution with 
the leading \order{\al^2} SM contribution with $\MH^{\SM} = 0$ and with 
the leading MSSM corrections originating from the $\Stop/\Sbot$ sector of
\order{\al} and \order{\al\als}. The numerical effect of the new
contribution exceeds the one of the leading QCD correction of 
\order{\al\als} in the scalar quark sector for $\msusy \gsim 500 \gev$.
It is always larger than the leading \order{\al^2} SM contribution with
$\MH^{\SM} = 0$, reaching approximately twice its value for small $\tb$
and moderate $\MA$.

The numerical effect of the new contribution on the precision
observables $\MW$ and $\sweff$ is reletively small, up to $-2 \mev$ for
$\MW$ and $+1 \times 10^{-5}$ for $\sweff$. It should be noted, however,
that the \order{\al^2} SM result with $\MH^{\SM} = 0$, to which the new
result corresponds, underestimates the result with realistic values of
$\MH^{\SM}$ by about one order of magnitude. A similar behavior can also
be expected for the MSSM corrections. An extension of our present result
to the case of non-zero values of the lightest $\cp$-even Higgs boson
mass will be undertaken in a forthcoming publication.


\newpage
\Acknowledgments
S.H. thanks the
organizers of ``RADCOR2000'' for 
the invitation and the inspiring
and constructive atmosphere at the workshop.


\end{document}